\documentclass[twocolumn,showpacs,superscriptaddress,altaffilletter]{revtex4}

\usepackage{graphicx}
\usepackage{dcolumn}
\usepackage{amsmath}

\def\kevc1{\ifmmode\mathrm{\ keV/{\mit c}}
          \else$\mathrm{\ keV/{\mit c}}$\fi}
\def\Mevc1{\ifmmode\mathrm{\ MeV/{\mit c}}
          \else$\mathrm{\ MeV/{\mit c}}$\fi}
\def\gevc1{\ifmmode\mathrm{\ GeV/{\mit c}}
          \else$\mathrm{\ GeV/{\mit c}}$\fi}
\def\kevc2{\ifmmode\mathrm{\ keV/{\mit c}^2}
          \else$\mathrm{\ keV/{\mit c}^2}$\fi}
\def\Mevc2{\ifmmode\mathrm{\ MeV/{\mit c}^2}
          \else$\mathrm{\ MeV/{\mit c}^2}$\fi}
\def\Gevc2{\ifmmode\mathrm{\ GeV/{\mit c}^2}
          \else$\mathrm{\ GeV/{\mit c}^2}$\fi}
\def\Gev2c2{\ifmmode\mathrm{\ GeV^2/{\mit c}^2}
          \else$\mathrm{\ GeV^2/{\mit c}^2}$\fi}

\def\ubar{\ifmmode\mathrm{\overline {u}}
          \else$\mathrm{\overline{u}}$\fi}
\def\dbar{\ifmmode\mathrm{\overline {d}}
          \else$\mathrm{\overline{d}}$\fi}
\def\sbar{\ifmmode\mathrm{\overline {s}}
          \else$\mathrm{\overline{s}}$\fi}
\def\cbar{\ifmmode\mathrm{\overline {c}}
          \else$\mathrm{\overline{c}}$\fi}
\def\bbar{\ifmmode\mathrm{\overline {b}}
          \else$\mathrm{\overline{b}}$\fi}
\def\tbar{\ifmmode\mathrm{\overline {t}}
          \else$\mathrm{\overline{t}}$\fi}
\def\qbar{\ifmmode\mathrm{\overline {q}}
          \else$\mathrm{\overline{q}}$\fi}

\def\uq{\ifmmode\mathrm{u}
          \else$\mathrm{u}$\fi}
\def\dq{\ifmmode\mathrm{d}
          \else$\mathrm{d}$\fi}
\def\sq{\ifmmode\mathrm{s}
          \else$\mathrm{s}$\fi}
\def\cq{\ifmmode\mathrm{c}
          \else$\mathrm{c}$\fi}
\def\bq{\ifmmode\mathrm{b}
          \else$\mathrm{b}$\fi}
\def\tq{\ifmmode\mathrm{t}
          \else$\mathrm{t}$\fi}
\def\qq{\ifmmode\mathrm{q}
          \else$\mathrm{q}$\fi}

 \def\Pgg{\ifmmode\mathrm{\gamma}
          \else$\mathrm{\gamma}$\fi}
 \def\PW{\ifmmode\mathrm{W}
         \else$\mathrm{W }$\fi}
 \def\PWp{\ifmmode\mathrm{W^+}
          \else$\mathrm{W^+}$\fi}
 \def\PWpm{\ifmmode\mathrm{W^{\pm}}
          \else$\mathrm{W^{\pm}}$\fi}
 \def\PWm{\ifmmode\mathrm{W^-}
          \else$\mathrm{W^-}$\fi}
 \def\PZz{\ifmmode\mathrm{Z^0}
          \else$\mathrm{Z^0}$\fi}
 \def\PHz{\ifmmode\mathrm{H^0}
          \else$\mathrm{H^0}$\fi}
 \def\PHpm{\ifmmode\mathrm{H^{\pm}}
           \else$\mathrm{H^{\pm}}$\fi}
 \def\PWR{\ifmmode\mathrm{W_R}
          \else$\mathrm{W_R}$\fi}
 \def\PWpr{\ifmmode\mathrm{W^{\prime}}
           \else$\mathrm{W^{\prime}}$\fi}
 \def\PZLR{\ifmmode\mathrm{Z_{LR}}
           \else$\mathrm{Z_{LR}}$\fi}
 \def\PZgc{\ifmmode\mathrm{Z_{\chi}}
           \else$\mathrm{Z_{\chi}}$\fi}
 \def\PZgy{\ifmmode\mathrm{Z_{\psi}}
           \else$\mathrm{Z_{\psi}}$\fi}
 \def\PZge{\ifmmode\mathrm{Z_{\eta}}
           \else$\mathrm{Z_{\eta}}$\fi}
 \def\PZi{\ifmmode\mathrm{Z_1}
          \else$\mathrm{Z_1}$\fi}
 \def\PAz{\ifmmode\mathrm{A^0}
          \else$\mathrm{A^0}$\fi}
 \def\Pgne{\ifmmode\mathrm{\nu_{e}}
           \else$\mathrm{\nu_{e}}$\fi}
 \def\Pagne{\ifmmode\mathrm{\overline{\nu_{e}}}
            \else$\mathrm{\overline{\nu_{e}}}$\fi}
 \def\Pgngm{\ifmmode\mathrm{\nu_{\mu}}
            \else$\mathrm{\nu_{\mu}}$\fi}
 \def\Pagngm{\ifmmode\mathrm{\overline{\nu}_{\mu}}
             \else$\mathrm{\overline{\nu}_{\mu}}$\fi}
 \def\Pgngt{\ifmmode\mathrm{\nu_{\tau}}
            \else$\mathrm{\nu_{\tau}}$\fi}
 \def\Pagngt{\ifmmode\mathrm{\overline{\nu}_{\tau}}
             \else$\mathrm{\overline{\nu}_{\tau}}$\fi}
 \def\Pe{\ifmmode\mathrm{e}
         \else$\mathrm{e}$\fi}
 \def\Pep{\ifmmode\mathrm{e^+}
          \else$\mathrm{e^+}$\fi}
 \def\Pem{\ifmmode\mathrm{e^-}
          \else$\mathrm{e^-}$\fi}
 \def\Pgm{\ifmmode\mathrm{\mu}
          \else$\mathrm{\mu}$\fi}
 \def\Pgmm{\ifmmode\mathrm{\mu^-}
           \else$\mathrm{\mu^-}$\fi}
 \def\Pgmp{\ifmmode\mathrm{\mu^+}
           \else$\mathrm{\mu^+}$\fi}
 \def\Pgt{\ifmmode\mathrm{\tau}
          \else$\mathrm{\tau}$\fi}
 \def\PLpm{\ifmmode\mathrm{L^{\pm}}
           \else$\mathrm{L^{\pm}}$\fi}
 \def\PLz{\ifmmode\mathrm{L^0}
          \else$\mathrm{L^0}$\fi}
 \def\PEz{\ifmmode\mathrm{E^0}
          \else$\mathrm{E^0}$\fi}
 \def\Pgp{\ifmmode\mathrm{\pi}
          \else$\mathrm{\pi }$\fi}
 \def\Pgpm{\ifmmode\mathrm{\pi^-}
           \else$\mathrm{\pi^-}$\fi}
 \def\Pgpp{\ifmmode\mathrm{\pi^+}
           \else$\mathrm{\pi^+}$\fi}
 \def\Pgppm{\ifmmode\mathrm{\pi^{\pm }}
            \else$\mathrm{\pi^{\pm }}$\fi}
 \def\Pgpz{\ifmmode\mathrm{\pi^0}
           \else$\mathrm{\pi^0 }$\fi}
 \def\Pgh{\ifmmode\mathrm{\eta}
          \else$\mathrm{\eta }$\fi}
 \def\Pgr{\ifmmode\mathrm{\rho(770)}
          \else$\mathrm{\rho(770)}$\fi}
 \def\Pgo{\ifmmode\mathrm{\omega(783)}
          \else$\mathrm{\omega(783)}$\fi}
 \def\Pghpr{\ifmmode\mathrm{\eta^{\prime}(958)}
            \else$\mathrm{\eta^{\prime}(958)}$\fi}
 \def\Pfz{\ifmmode\mathrm{f_0(980)}
          \else$\mathrm{f_0(980)}$\fi}
 \def\Paz{\ifmmode\mathrm{a_0(980)}
          \else$\mathrm{a_0(980)}$\fi}
 \def\Pgf{\ifmmode\mathrm{\phi(1020)}
          \else$\mathrm{\phi(1020)}$\fi}
 \def\Phia{\ifmmode\mathrm{h_1(1170)}
           \else$\mathrm{h_1(1170)}$\fi}
 \def\Pbi{\ifmmode\mathrm{b_1(1235)}
          \else$\mathrm{b_1(1235)}$\fi}
 \def\Pai{\ifmmode\mathrm{a_1(1260)}
          \else$\mathrm{a_1(1260)}$\fi}
 \def\Pfii{\ifmmode\mathrm{f_2(1270)}
           \else$\mathrm{f_2(1270)}$\fi}
 \def\Pfi{\ifmmode\mathrm{f_1(1285)}
          \else$\mathrm{f_1(1285)}$\fi}
 \def\Pgha{\ifmmode\mathrm{\eta(1295)}
           \else$\mathrm{\eta(1295)}$\fi}
 \def\Pgpa{\ifmmode\mathrm{\pi(1300)}
           \else$\mathrm{\pi(1300)}$\fi}
 \def\Paii{\ifmmode\mathrm{a_2(1320)}
           \else$\mathrm{a_2(1320)}$\fi}
 \def\Pgoa{\ifmmode\mathrm{\omega(1390)}
           \else$\mathrm{\omega(1390)}$\fi}
 \def\Pfza{\ifmmode\mathrm{f_0(1400)}
           \else$\mathrm{f_0(1400)}$\fi}
 \def\Pfia{\ifmmode\mathrm{f_1 (1390)}
           \else$\mathrm{f_1 (1390)}$\fi}
 \def\Pghb{\ifmmode\mathrm{\eta(1440)}
           \else$\mathrm{\eta(1440)}$\fi}
 \def\Pgra{\ifmmode\mathrm{\rho(1450)}
           \else$\mathrm{\rho(1450)}$\fi}
 \def\Pfib{\ifmmode\mathrm{f_1(1510)}
           \else$\mathrm{f_1(1510)}$\fi}
 \def\Pfiipr{\ifmmode\mathrm{f^{\prime}_2(1525)}
             \else$\mathrm{f^{\prime}_2(1525)}$\fi}
 \def\Pfzb{\ifmmode\mathrm{f_0(1590)}
           \else$\mathrm{f_0(1590)}$\fi}
 \def\Pgob{\ifmmode\mathrm{\omega(1600)}
           \else$\mathrm{\omega(1600)}$\fi}
 \def\Pgoiii{\ifmmode\mathrm{\omega_3(1670)}
             \else$\mathrm{\omega_3(1670)}$\fi}
 \def\Pgpii{\ifmmode\mathrm{\pi_2(1670)}
            \else$\mathrm{\pi_2(1670)}$\fi}
 \def\Pgfa{\ifmmode\mathrm{\phi(1680)}
           \else$\mathrm{\phi(1680)}$\fi}
 \def\Pgriii{\ifmmode\mathrm{\rho_3(1690)}
             \else$\mathrm{\rho_3(1690)}$\fi}
 \def\Pgrb{\ifmmode\mathrm{\rho(1700)}
           \else$\mathrm{\rho(1700)}$\fi}
 \def\Pfiia{\ifmmode\mathrm{f_2(1720)}
            \else$\mathrm{f_2(1720)}$\fi}
 \def\Pgfiii{\ifmmode\mathrm{\phi_3(1850)}
             \else$\mathrm{\phi_3(1850)}$\fi}
 \def\Pfiib{\ifmmode\mathrm{f_2(2010)}
            \else$\mathrm{f_2(2010)}$\fi}
 \def\Pfiv{\ifmmode\mathrm{f_4(2050)}
           \else$\mathrm{f_4(2050)}$\fi}
 \def\Pfiic{\ifmmode\mathrm{f_2(2300)}
            \else$\mathrm{f_2(2300)}$\fi}
 \def\Pfiid{\ifmmode\mathrm{f_2(2340)}
            \else$\mathrm{f_2(2340)}$\fi}
 \def\PK{\ifmmode\mathrm{K}
         \else$\mathrm{K}$\fi}
 \def\PKpm{\ifmmode\mathrm{K^{\pm}}
           \else$\mathrm{K^{\pm}}$\fi}
 \def\PKp{\ifmmode\mathrm{K^+}
          \else$\mathrm{K^+}$\fi}
 \def\PKm{\ifmmode\mathrm{K^-}
          \else$\mathrm{K^-}$\fi}
 \def\PKz{\ifmmode\mathrm{K^0}
          \else$\mathrm{K^0}$\fi}
 \def\PaKz{\ifmmode\mathrm{\overline{K^0}}
           \else$\mathrm{\overline{K^0}}$\fi}
 \def\PKgmiii{\ifmmode\mathrm{K_{\mu 3}}
              \else$\mathrm{K_{\mu 3}}$\fi}
 \def\PKeiii{\ifmmode\mathrm{K_{\rm e3}}
             \else$\mathrm{K_{\rm e3}}$\fi}
 \def\PKzS{\ifmmode\mathrm{K^0_{\rm S}}
           \else$\mathrm{K^0_{\rm S}}$\fi}
 \def\PKzL{\ifmmode\mathrm{K^0_{\rm L}}
           \else$\mathrm{K^0_{\rm L}}$\fi}
 \def\PKzgmiii{\ifmmode\mathrm{K^0_{\mu 3}}
               \else$\mathrm{K^0_{\mu 3}}$\fi}
 \def\PKzeiii{\ifmmode\mathrm{K^0_{{\rm e}3}}
              \else$\mathrm{K^0_{{\rm e}3}}$\fi}
 \def\PKst{\ifmmode\mathrm{K^{\ast}(892)}
           \else$\mathrm{K^{\ast}(892)}$\fi}
 \def\PKi{\ifmmode\mathrm{K_1(1270)}
          \else$\mathrm{K_1(1270)}$\fi}
 \def\PKsta{\ifmmode\mathrm{K^{\ast}(1370)}
            \else$\mathrm{K^{\ast}(1370)}$\fi}
 \def\PKia{\ifmmode\mathrm{K_1(1400)}
           \else$\mathrm{K_1(1400)}$\fi}
 \def\PKstz{\ifmmode\mathrm{K^{\ast}_0(1430)}
            \else$\mathrm{K^{\ast}_0(1430)}$\fi}
 \def\PKstii{\ifmmode\mathrm{K^{\ast}_2(1430)}
             \else$\mathrm{K^{\ast}_2(1430)}$\fi}
 \def\PKstb{\ifmmode\mathrm{K^{\ast}(1680)}
            \else$\mathrm{K^{\ast}(1680)}$\fi}
 \def\PKii{\ifmmode\mathrm{K_2(1770)}
           \else$\mathrm{K_2(1770)}$\fi}
 \def\PKstiii{\ifmmode\mathrm{K^{\ast}_3(1780)}
              \else$\mathrm{K^{\ast}_3(1780)}$\fi}
 \def\PKstiv{\ifmmode\mathrm{K^{\ast}_4(2045)}
             \else$\mathrm{K^{\ast}_4(2045)}$\fi}
 \def\PD{\ifmmode\mathrm{D}
           \else$\mathrm{D}$\fi}
 \def\PaD{\ifmmode\mathrm{\overline{ D}}
          \else${\mathrm{\overline D}}$\fi}
 \def\PDpm{\ifmmode\mathrm{D^{\pm}}
           \else$\mathrm{D^{\pm}}$\fi}
 \def\PDm{\ifmmode\mathrm{D^-}
          \else$\mathrm{D^-}$\fi}
 \def\PDp{\ifmmode\mathrm{D^+}
          \else$\mathrm{D^+}$\fi}
 \def\PDz{\ifmmode\mathrm{D^0}
          \else$\mathrm{D^0}$\fi}
 \def\PaDz{\ifmmode\mathrm{\overline{D^0}}
           \else$\mathrm{\overline{D^0}}$\fi}
 \def\PDstpm{\ifmmode{\mathrm{D}^{\ast}(2010)^{\pm}}
             \else$\mathrm{D}^{\ast}(2010)^{\pm}$\fi}
 \def\PDstp{\ifmmode{\mathrm{D}^{\ast+}}
             \else$\mathrm{D}^{\ast+}$\fi}
 \def\PDst{\ifmmode{\mathrm{D}^{\ast}}
             \else$\mathrm{D}^{\ast}$\fi}
 \def\PDstz{\ifmmode{\mathrm{D}^{\ast}(2010)^0}
            \else$\mathrm{D}^{\ast}(2010)^0$\fi}
 \def\PDiz{\ifmmode{\mathrm{D}_{1}(2420)^0}
           \else$\mathrm{D}_{1}(2420)^0$\fi}
 \def\PDstiiz{\ifmmode{\mathrm{D}^{\ast}_{2}(2460)^0}
              \else$\mathrm{D}^{\ast}_{2}(2460)^0$\fi}
 \def\PsDp{\ifmmode\mathrm{D_{s}^+}
           \else$\mathrm{D_{s}^+}$\fi}
 \def\PsDm{\ifmmode\mathrm{D_{s}^-}
           \else$\mathrm{D_{s}^-}$\fi}
 \def\PsDpm{\ifmmode\mathrm{D_{s}^{\pm}}
           \else$\mathrm{D_{s}^{\pm}}$\fi}
 \def\PsDst{\ifmmode\mathrm{D_{s}^{\ast}}
            \else$\mathrm{D_{s}^{\ast}}$\fi}
 \def\PsDipm{\ifmmode\mathrm{D_{s1}(2536)^{\pm}}
           \else$\mathrm{D_{s1}(2536)^{\pm}}$\fi}
 \def\PB{\ifmmode{\mathrm{B}}
          \else$\mathrm{B}$\fi}
 \def\PBp{\ifmmode{\mathrm{B}^{+}}
           \else$\mathrm{B}^{+}$\fi}
 \def\PBm{\ifmmode{\mathrm{B}^{-}}
           \else$\mathrm{B}^{-}$\fi}
 \def\PBpm{\ifmmode{\mathrm{B}^{\pm}}
            \else$\mathrm{B}^{\pm}$\fi}
 \def\PBz{\ifmmode{\mathrm{B}^0}
           \else$\mathrm{B}^0$\fi}
 \def\PbgL{\ifmmode{\mathrm{\Lambda}_b}
           \else$\mathrm{\Lambda}_b$\fi}
 \def\Pcgh{\ifmmode\mathrm{{\eta}_{c}(1S)}
           \else$\mathrm{{\eta}_{c}(1S)}$\fi}
 \def\PJgyy{\ifmmode\mathrm{J /\psi}
           \else$\mathrm{J /\psi}$\fi}
 \def\PJgy{\ifmmode\mathrm{J /\psi(1S)}
           \else$\mathrm{J /\psi(1S)}$\fi}
 \def\Pcgcz{\ifmmode\mathrm{{\chi}_{c0}(1P)}
            \else$\mathrm{{\chi}_{c0}(1P)}$\fi}
 \def\Pcgci{\ifmmode\mathrm{{\chi}_{c1}(1P)}
            \else$\mathrm{{\chi}_{c1}(1P)}$\fi}
 \def\Pcgcii{\ifmmode\mathrm{{\chi}_{c2}(1P)}
             \else$\mathrm{{\chi}_{c2}(1P)}$\fi}
 \def\Pgy{\ifmmode\mathrm{\psi(2S)}
          \else$\mathrm{\psi(2S)}$\fi}
 \def\Pgya{\ifmmode\mathrm{\psi(3770)}
           \else$\mathrm{\psi(3770)}$\fi}
 \def\Pgyb{\ifmmode\mathrm{\psi(4040)}
           \else$\mathrm{\psi(4040)}$\fi}
 \def\Pgyc{\ifmmode\mathrm{\psi(4160)}
           \else$\mathrm{\psi(4160)}$\fi}
 \def\Pgyd{\ifmmode\mathrm{\psi(4415)}
           \else$\mathrm{\psi(4415)}$\fi}
 \def\PgU{\ifmmode\mathrm{\Upsilon(1S)}
          \else$\mathrm{\Upsilon(1S)}$\fi}
 \def\Pbgcz{\ifmmode\mathrm{{\chi}_{b0}(1P)}
            \else$\mathrm{{\chi}_{b0}(1P)}$\fi}
 \def\Pbgci{\ifmmode\mathrm{{\chi}_{b1}(1P)}
            \else$\mathrm{{\chi}_{b1}(1P)}$\fi}
 \def\Pbgcii{\ifmmode\mathrm{{\chi}_{b2}(1P)}
             \else$\mathrm{{\chi}_{b2}(1P)}$\fi}
 \def\PgUa{\ifmmode\mathrm{\Upsilon(2S)}
           \else$\mathrm{\Upsilon(2S)}$\fi}
 \def\Pbgcza{\ifmmode\mathrm{{\chi}_{b0}(2P)}
             \else$\mathrm{{\chi}_{b0}(2P)}$\fi}
 \def\Pbgcia{\ifmmode\mathrm{{\chi}_{b1}(2P)}
             \else$\mathrm{{\chi}_{b1}(2P)}$\fi}
 \def\Pbgciia{\ifmmode\mathrm{{\chi}_{b2}(2P)}
              \else$\mathrm{{\chi}_{b2}(2P)}$\fi}
 \def\PgUb{\ifmmode\mathrm{\Upsilon(3S)}
           \else$\mathrm{\Upsilon(3S)}$\fi}
 \def\PgUc{\ifmmode\mathrm{\Upsilon(4S)}
           \else$\mathrm{\Upsilon(4S)}$\fi}
 \def\PgUd{\ifmmode\mathrm{\Upsilon(10860)}
           \else$\mathrm{\Upsilon(10860)}$\fi}
 \def\PgUe{\ifmmode\mathrm{\Upsilon(11020)}
           \else$\mathrm{\Upsilon(11020)}$\fi}
 \def\Pp{\ifmmode\mathrm{p}
         \else$\mathrm{p}$\fi}
 \def\Pap{\ifmmode\mathrm{\overline{p}}
         \else$\mathrm{\overline{p}}$\fi}
 \def\Pn{\ifmmode\mathrm{n}
         \else$\mathrm{n}$\fi}
 \def\PNa{\ifmmode\mathrm{N(1440)P_{11}}
          \else$\mathrm{N(1440)P_{11}}$\fi}
 \def\PNb{\ifmmode\mathrm{N(1520)D_{13}}
          \else$\mathrm{N(1520)D_{13}}$\fi}
 \def\PNc{\ifmmode\mathrm{N(1535)S_{11}}
          \else$\mathrm{N(1535)S_{11}}$\fi}
 \def\PNd{\ifmmode\mathrm{N(1650)S_{11}}
          \else$\mathrm{N(1650)S_{11}}$\fi}
 \def\PNe{\ifmmode\mathrm{N(1675)D_{15}}
          \else$\mathrm{N(1675)D_{15}}$\fi}
 \def\PNf{\ifmmode\mathrm{N(1680)F_{15}}
          \else$\mathrm{N(1680)F_{15}}$\fi}
 \def\PNg{\ifmmode\mathrm{N(1700)D_{13}}
          \else$\mathrm{N(1700)D_{13}}$\fi}
 \def\PNh{\ifmmode\mathrm{N(1710)P_{11}}
          \else$\mathrm{N(1710)P_{11}}$\fi}
 \def\PNi{\ifmmode\mathrm{N(1720)P_{13}}
          \else$\mathrm{N(1720)P_{13}}$\fi}
 \def\PNj{\ifmmode\mathrm{N(2190)G_{17}}
          \else$\mathrm{N(2190)G_{17}}$\fi}
 \def\PNk{\ifmmode\mathrm{N(2220)H_{19}}
          \else$\mathrm{N(2220)H_{19}}$\fi}
 \def\PNl{\ifmmode\mathrm{N(2250)G_{19}}
          \else$\mathrm{N(2250)G_{19}}$\fi}
 \def\PNm{\ifmmode\mathrm{N(2600)I_{1,11}}
          \else$\mathrm{N(2600)I_{1,11}}$\fi}
 \def\PgDa{\ifmmode\mathrm{\Delta(1232)P_{33}}
           \else$\mathrm{\Delta(1232)P_{33}}$\fi}
 \def\PgDb{\ifmmode\mathrm{\Delta(1620)S_{31}}
           \else$\mathrm{\Delta(1620)S_{31}}$\fi}
 \def\PgDc{\ifmmode\mathrm{\Delta(1700)D_{33}}
           \else$\mathrm{\Delta(1700)D_{33}}$\fi}
 \def\PgDd{\ifmmode\mathrm{\Delta(1900)S_{31}}
           \else$\mathrm{\Delta(1900)S_{31}}$\fi}
 \def\PgDe{\ifmmode\mathrm{\Delta(1905)F_{35}}
           \else$\mathrm{\Delta(1905)F_{35}}$\fi}
 \def\PgDf{\ifmmode\mathrm{\Delta(1910)P_{31}}
           \else$\mathrm{\Delta(1910)P_{31}}$\fi}
 \def\PgDh{\ifmmode\mathrm{\Delta(1920)P_{33}}
           \else$\mathrm{\Delta(1920)P_{33}}$\fi}
 \def\PgDi{\ifmmode\mathrm{\Delta(1930)D_{35}}
           \else$\mathrm{\Delta(1930)D_{35}}$\fi}
 \def\PgDj{\ifmmode\mathrm{\Delta(1950)F_{37}}
           \else$\mathrm{\Delta(1950)F_{37}}$\fi}
 \def\PgDk{\ifmmode\mathrm{\Delta(2420)H_{3,11}}
           \else$\mathrm{\Delta(2420)H_{3,11}}$\fi}
 \def\PgDpp{\ifmmode\mathrm{\Delta^{++}}
           \else$\mathrm{\Delta^{++}}$\fi}
 \def\PgL{\ifmmode\mathrm{\Lambda}
          \else$\mathrm{\Lambda}$\fi}
 \def\PagL{\ifmmode\mathrm{\overline{\Lambda}}
            \else$\mathrm{\overline{\Lambda}}$\fi}
 \def\PgLa{\ifmmode\mathrm{\Lambda(1405) S_{01}}
           \else$\mathrm{\Lambda(1405) S_{01}}$\fi}
 \def\PgLb{\ifmmode\mathrm{\Lambda(1520) D_{03}}
           \else$\mathrm{\Lambda(1520) D_{03}}$\fi}
 \def\PgLc{\ifmmode\mathrm{\Lambda(1600) P_{01}}
           \else$\mathrm{\Lambda(1600) P_{01}}$\fi}
 \def\PgLd{\ifmmode\mathrm{\Lambda(1670) S_{01}}
           \else$\mathrm{\Lambda(1670) S_{01}}$\fi}
 \def\PgLe{\ifmmode\mathrm{\Lambda(1690) D_{03}}
           \else$\mathrm{\Lambda(1690) D_{03}}$\fi}
 \def\PgLf{\ifmmode\mathrm{\Lambda(1800) S_{01}}
           \else$\mathrm{\Lambda(1800) S_{01}}$\fi}
 \def\PgLg{\ifmmode\mathrm{\Lambda(1810) P_{01}}
           \else$\mathrm{\Lambda(1810) P_{01}}$\fi}
 \def\PgLh{\ifmmode\mathrm{\Lambda(1820) F_{05}}
           \else$\mathrm{\Lambda(1820) F_{05}}$\fi}
 \def\PgLi{\ifmmode\mathrm{\Lambda(1830) D_{05}}
           \else$\mathrm{\Lambda(1830) D_{05}}$\fi}
 \def\PgLj{\ifmmode\mathrm{\Lambda(1890) P_{03}}
           \else$\mathrm{\Lambda(1890) P_{03}}$\fi}
 \def\PgLk{\ifmmode\mathrm{\Lambda(2100) G_{07}}
           \else$\mathrm{\Lambda(2100) G_{07}}$\fi}
 \def\PgLl{\ifmmode\mathrm{\Lambda(2110) F_{05}}
           \else$\mathrm{\Lambda(2110) F_{05}}$\fi}
 \def\PgLm{\ifmmode\mathrm{\Lambda(2350) H_{09}}
           \else$\mathrm{\Lambda(2350) H_{09}}$\fi}
 \def\PgS{\ifmmode{\rm \Sigma}
           \else${\rm \Sigma}$\fi}
 \def\PgSp{\ifmmode\mathrm{\Sigma^+}
           \else$\mathrm{\Sigma^+}$\fi}
 \def\PgSz{\ifmmode\mathrm{\Sigma^0}
           \else$\mathrm{\Sigma^0}$\fi}
 \def\PgSm{\ifmmode\mathrm{\Sigma^-}
           \else$\mathrm{\Sigma^-}$\fi}
 \def\PgSpm{\ifmmode\mathrm{\Sigma^{\pm}}
           \else$\mathrm{\Sigma^{\pm}}$\fi}
 \def\PgSa{\ifmmode\mathrm{\Sigma(1385) P_{13}}
           \else$\mathrm{\Sigma(1385) P_{13}}$\fi}
 \def\PgSb{\ifmmode\mathrm{\Sigma(1660) P_{11}}
           \else$\mathrm{\Sigma(1660) P_{11}}$\fi}
 \def\PgSc{\ifmmode\mathrm{\Sigma(1670) D_{13}}
           \else$\mathrm{\Sigma(1670) D_{13}}$\fi}
 \def\PgSd{\ifmmode\mathrm{\Sigma(1750) S_{11}}
           \else$\mathrm{\Sigma(1750) S_{11}}$\fi}
 \def\PgSe{\ifmmode\mathrm{\Sigma(1775) D_{15}}
           \else$\mathrm{\Sigma(1775) D_{15}}$\fi}
 \def\PgSf{\ifmmode\mathrm{\Sigma(1915) F_{15}}
           \else$\mathrm{\Sigma(1915) F_{15}}$\fi}
 \def\PgSg{\ifmmode\mathrm{\Sigma(1940) D_{13}}
           \else$\mathrm{\Sigma(1940) D_{13}}$\fi}
 \def\PgSh{\ifmmode\mathrm{\Sigma(2030) F_{17}}
           \else$\mathrm{\Sigma(2030) F_{17}}$\fi}
 \def\PgSi{\ifmmode\mathrm{\Sigma(2050)}
           \else$\mathrm{\Sigma(2050)}$\fi}
 \def\PgXz{\ifmmode\mathrm{\Xi^0}
           \else$\mathrm{\Xi^0}$\fi}
 \def\PgXm{\ifmmode\mathrm{\Xi^-}
           \else$\mathrm{\Xi^-}$\fi}
 \def\PgXa{\ifmmode\mathrm{\Xi(1530)}
           \else$\mathrm{\Xi(1530)}$\fi}
 \def\PgXas{\ifmmode\mathrm{\Xi(1530)P_{13}}
           \else$\mathrm{\Xi(1530)P_{13}}$\fi}
 \def\PgXb{\ifmmode\mathrm{\Xi(1690)}
           \else$\mathrm{\Xi(1690)}$\fi}
 \def\PgXbb{\ifmmode\mathrm{\Xi(1620)}
           \else$\mathrm{\Xi(1620)}$\fi}
 \def\PgXc{\ifmmode\mathrm{\Xi(1820)D_{13}}
           \else$\mathrm{\Xi(1820)D_{13}}$\fi}
 \def\PgXcs{\ifmmode\mathrm{\Xi(1820)}
           \else$\mathrm{\Xi(1820)}$\fi}
 \def\PgXd{\ifmmode\mathrm{\Xi(1950)}
           \else$\mathrm{\Xi(1950)}$\fi}
 \def\PgXe{\ifmmode\mathrm{\Xi(2030)}
           \else$\mathrm{\Xi(2030)}$\fi}
 \def\PgOm{\ifmmode\mathrm{\Omega^-}
           \else$\mathrm{\Omega^-}$\fi}
 \def\PgO{\ifmmode\mathrm{\Omega}
           \else$\mathrm{\Omega}$\fi}
 \def\PgOma{\ifmmode\mathrm{\Omega(2250)^-}
            \else$\mathrm{\Omega(2250)^-}$\fi}
 \def\PcgL{\ifmmode\mathrm{\Lambda_c}
            \else$\mathrm{\Lambda_c}$\fi}
 \def\PacgL{\ifmmode\mathrm{\overline{\Lambda}_c}
            \else$\mathrm{\overline{\Lambda}_c}$\fi}
 \def\PcgLp{\ifmmode\mathrm{\Lambda_c^+}
            \else$\mathrm{\Lambda_c^+}$\fi}
 \def\PcgLm{\ifmmode{\rm \Lambda_c^-}
            \else${\rm \Lambda_c^-}$\fi}
 \def\PcgX{\ifmmode\mathrm{\Xi_c}
            \else$\mathrm{\Xi_c}$\fi}
 \def\PcgXz{\ifmmode\mathrm{\Xi_c^0}
            \else$\mathrm{\Xi_c^0}$\fi}
 \def\PcgXp{\ifmmode\mathrm{\Xi_c^+}
            \else$\mathrm{\Xi_c^+}$\fi}
 \def\PcgS{\ifmmode\mathrm{\Sigma_c}
           \else$\mathrm{\Sigma_c}$\fi}
 \def\PcgSz{\ifmmode\mathrm{\Sigma_c^0}
           \else$\mathrm{\Sigma_c^0}$\fi}
 \def\PcgSp{\ifmmode\mathrm{\Sigma_c^+}
           \else$\mathrm{\Sigma_c^+}$\fi}
 \def\PcgSpp{\ifmmode\mathrm{\Sigma_c^{++}}
           \else$\mathrm{\Sigma_c^{++}}$\fi}
 \def\PcgO{\ifmmode{\mathrm \Omega_c}
           \else${\mathrm \Omega_c}$\fi}
 \def\PcgOz{\ifmmode{\mathrm \Omega_c^{0}}
           \else${\mathrm \Omega_c^{0}}$\fi}
 \def\PSgg{\ifmmode\mathrm{\tilde{\gamma}}
           \else$\mathrm{\tilde{\gamma}}$\fi}
 \def\PSgxz{\ifmmode\mathrm{\tilde{\chi}^0_i}
            \else$\mathrm{\tilde{\chi}^0_i}$\fi}
 \def\PSZz{\ifmmode\mathrm{\tilde{Z}^0}
           \else$\mathrm{\tilde{Z}^0}$\fi}
 \def\PSHz{\ifmmode\mathrm{\tilde{H}^0_j}
           \else$\mathrm{\tilde{H}^0_j}$\fi}
 \def\PSgxpm{\ifmmode\mathrm{\tilde{\chi}^{\pm_i}}
             \else$\mathrm{\tilde{\chi}^{\pm_i}}$\fi}
 \def\PSWpm{\ifmmode\mathrm{\tilde{W}^{\pm}}
            \else$\mathrm{\tilde{W}^{\pm}}$\fi}
 \def\PSHpm{\ifmmode\mathrm{\tilde{H}^{\pm_j}}
            \else$\mathrm{\tilde{H}^{\pm_j}}$\fi}
 \def\PSgn{\ifmmode\mathrm{\tilde{\nu}}
           \else$\mathrm{\tilde{\nu}}$\fi}
 \def\PSe{\ifmmode\mathrm{\tilde{e}}
          \else$\mathrm{\tilde{e}}$\fi}
 \def\PSgm{\ifmmode\mathrm{\tilde{\mu}}
           \else$\mathrm{\tilde{\mu}}$\fi}
 \def\PSgt{\ifmmode\mathrm{\tilde{\tau}}
           \else$\mathrm{\tilde{\tau}}$\fi}
 \def\PSq{\ifmmode\mathrm{\tilde{q}}
          \else$\mathrm{\tilde{q}}$\fi}
 \def\PSg{\ifmmode\mathrm{\tilde{g}}
          \else$\mathrm{\tilde{g}}$\fi}

\def\ks{$K^0_s$}
\def\ksp{$K^0_s p$}
\def\la{$\Lambda$}
\def\tp{\ifmmode\mathrm{\Theta^+}
          \else$\mathrm{\Theta^+}$\fi}
\def\t1540{\ifmmode{\mathrm \Theta(1540)^+}
           \else${\mathrm \Theta(1540)^+}$\fi}
\def\xf{$x_F$}
\def\pt{$p_t$}

\def\insta {University of Bristol, Bristol, United Kingdom}
\def\instb {CERN, CH-1211 Gen\`eve 23, Switzerland}
\def\instc {Genoa University/INFN, Dipartimento di Fisica,I-16146 Genova, Italy}
\def\instd {Grenoble ISN, F-38026 Grenoble, France}
\def\inste {Max-Planck-Institut f\"ur Kernphysik, Postfach 103980, D-69029 Heidelberg, Germany}
\def\instf {Universit\"at Heidelberg, Physikalisches Institut, D-69120 Heidelberg, Germany}
\def\instg {Universit\"at Mainz, Institut f\"ur Kernphysik, D-55099 Mainz, Germany}
\def\insth {Moscow Lebedev Physics Institute, RU-117924, Moscow, Russia}
\def\insti {University of Iowa, Iowa City, IA 52242, USA}
\def\instj {Rutgers University, Piscataway, New Jersey 08854, USA}
\def\instk {NIKHEF, 1009 D8 Amsterdam, The Netherlands}

\begin{document}

\preprint{xxx}

\title[Search for Thetaplus]
{Search for the pentaquark candidate \t1540\ in the hyperon beam
experiment WA89}

\author{M.I.~Adamovich}
\thanks{Deceased.}
\affiliation{\insth}
\author{Yu.A.~Alexandrov}
\thanks{Supported by the Deutsche Forschungsgemeinschaft,
           contract number436 RUS 113/465, and Russian Foundation for
           Basic Research under contract number RFFI 98-02-04096.}
\affiliation{\insth}
\author{S.P.~Baranov}
\affiliation{\insth}
\author{D.~Barberis}
\affiliation{\instc}
\author{M.~Beck}
\affiliation{\inste}
\author{C.~B\'erat}
\affiliation{\instd}
\author{W.~Beusch}
\affiliation{\instb}
\author{M.~Boss}
\affiliation{\instf}
\thanks{Supported by the Bundesministerium f\"ur Bildung, Wissenschaft,
Forschung und Technologie, Germany, under contract numbers
05HD515I and 06HD524I.}
\author{S.~Brons}
\altaffiliation{Present address: TRIUMF, Vancouver, B.C., Canada
V6T 2A3} \affiliation{\inste}
\author{W.~Br\"uckner}
\affiliation{\inste}
\author{M.~Bu\'enerd}
\affiliation{\instd}
\author{C.~Busch}
\affiliation{\instf}
\author{C.~B\"uscher}
\affiliation{\inste}
\author{F.~Charignon}
\affiliation{\instd}
\author{J.~Chauvin}
\affiliation{\instd}
\author{E.A.~Chudakov}
\altaffiliation{Present address: Thomas Jefferson Lab, Newport
News, VA 23606, USA.} \affiliation{\instf}
\author{U.~Dersch}
\affiliation{\inste}
\author{F.~Dropmann}
\affiliation{\inste}
\author{J.~Engelfried}
\altaffiliation{Present address: Institudo de Fisica, Universidad
San Luis Potosi, S.L.P. 78240, Mexico.} \affiliation{\instf}
\author{F.~Faller}
\altaffiliation{Present address: Fraunhofer Institut f\"ur
Solarenergiesysteme, D-79100 Freiburg, Germany.}
\affiliation{\instf}
\author{A.~Fournier}
\affiliation{\instd}
\author{S.G.~Gerassimov}
\altaffiliation{Present address: Fakult\"at f\"ur Physik,
Universit\"at Freiburg, Germany.} \affiliation{\inste}
\affiliation{\insth}
\author{M.~Godbersen}
\affiliation{\inste}
\author{P.~Grafstr\"om}
\affiliation{\instb}
\author{Th.~Haller}
\affiliation{\inste}
\author{M.~Heidrich}
\affiliation{\inste}
\author{E.~Hubbard}
\affiliation{\inste}
\author{R.B.~Hurst}
\affiliation{\instc}
\author{K.~K\"onigsmann}
\altaffiliation{Present address: Fakult\"at f\"ur Physik,
Universit\"at Freiburg, Germany.} \affiliation{\inste}
\author{I.~Konorov}
\altaffiliation{Present address: Technische Universit\"at
M\"unchen, Garching, Germany.} \affiliation{\inste}
\affiliation{\insth}
\author{N.~Keller}
\affiliation{\instf}
\author{K.~Martens}
\altaffiliation{Present address: Department of Physics and
Astronomy, SUNY at Stony Brook, NY 11794-3800, USA.}
\affiliation{\instf}
\author{Ph.~Martin}
\affiliation{\instd}
\author{S.~Masciocchi}
\altaffiliation{Present address: Max-Planck-Institut f\"ur Physik,
M\"unchen, Germany.} \affiliation{\inste}
\author{R.~Michaels}
\altaffiliation{Present address: Thomas Jefferson Lab, Newport
News, VA 23606, USA.} \affiliation{\inste}
\author{U.~M\"uller}
\affiliation{\instg}
\thanks{Supported by the Bundesministerium f\"ur Bildung, Wissenschaft,
Forschung und Technologie, Germany, under contract number 06MZ5265
and 06MZ177.}
\author{H.~Neeb}
\affiliation{\inste}
\author{D.~Newbold}
\affiliation{\insta}
\author{C.~Newsom}
\affiliation{\insti}
\author{S.~Paul}
\altaffiliation{Present address: Technische Universit\"at
M\"unchen, Garching, Germany.} \affiliation{\inste}
\author{J.~Pochodzalla}
\email[Contact person:]{pochodza@kph.uni-mainz.de}
\altaffiliation{present address: Universit\"at Mainz, Institut
f\"ur Kernphysik, D-55099 Mainz, Germany.} \affiliation{\inste}
\author{I.~Potashnikova}
\affiliation{\inste}
\author{B.~Povh}
\affiliation{\inste}
\author{R.~Ransome}
\affiliation{\instj}
\author{Z.~Ren}
\affiliation{\inste}
\author{M.~Rey-Campagnolle}
\altaffiliation {Present address: CERN, CH-1211 Gen\`eve 23,
Switzerland} \affiliation{\instd}
\author{G.~Rosner}
\altaffiliation {Present address: Dept. of Physics and Astronomy,
University of Glasgow, Glasgow G12 8QQ, United Kingdom}
\affiliation{\instg}
\author{L.~Rossi}
\affiliation{\instc}
\author{H.~Rudolph}
\affiliation{\instg}
\author{C.~Scheel}
\affiliation{\instk}
\author{L.~Schmitt}
\altaffiliation{Present address: Technische Universit\"at
M\"unchen, Garching, Germany.} \affiliation{\instg}
\author{H.-W.~Siebert}
\altaffiliation{Present address: Universit\"at Mainz, Institut
f\"ur Kernphysik, D-55099 Mainz, Germany.}\affiliation{\instf}
\author{A.~Simon}
\altaffiliation{Present address: Fakult\"at f\"ur Physik,
Universit\"at
 Freiburg, Germany.}
\affiliation{\instf}
\author{V.J.~Smith}
\altaffiliation{Supported by the UK PPARC} \noaffiliation
\affiliation{\insta}
\author{O.~Thilmann}
\affiliation{\instf}
\author{A.~Trombini}
\affiliation{\inste}
\author{E.~Vesin}
\affiliation{\instd}
\author{B.~Volkemer}
\affiliation{\instg}
\author{K.~Vorwalter}
\affiliation{\inste}
\author{Th.~Walcher}
\affiliation{\instg}
\author{G.~W\"alder}
\affiliation{\instf}
\author{R.~Werding}
\affiliation{\inste}
\author{E.~Wittmann}
\affiliation{\inste}
\author{M.V.~Zavertyaev}
\altaffiliation{Supported by the Deutsche Forschungsgemeinschaft,
contract number436 RUS 113/465, and Russian Foundation for Basic
Research under contract number RFFI 98-02-04096.}
\affiliation{\insth}

\collaboration{WA89 collaboration} \noaffiliation

\date{\today}

\begin{abstract}
We report on a high-statistics search for the \t1540\ resonance in
$\Sigma^-$-nucleus collisions at 340 \gevc1 . No evidence for this
resonance was found in our data sample which contains 13 millions
$K^0_s \rightarrow \pi^+\pi^- $ decays above background. For the
decay channel $\Theta^+ \rightarrow K^0_s p$ and the kinematic
range $x_F>$ 0.05 we find the production cross section to be
$BR(\Theta^+ \rightarrow K^0_s p)\cdot \sigma_0\, <$ 1.8 $\mu$b
per nucleon at 99\% CL.
\end{abstract}


\maketitle

\begin{figure}[t]
 \vspace{5.5cm}
\includegraphics{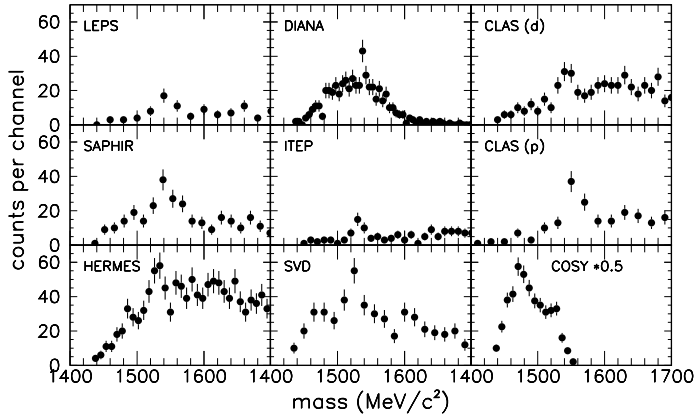}
 \caption{\it
      Summary of the first nine published observations of the {\t1540}
      resonance \cite{Theta:LEPS,Theta:DIANA,Theta:CLASd,Theta:SAPHIR,Theta:CLASp,
Theta:NEUTRINO,Theta:HERMES,Theta:SVD,Theta:COSY}.}
\label{fig:WA8902}
\end{figure}

During the last years twelve experimental groups have reported
evidence for a narrow baryonic resonance in the KN channel  at a
mass of about 1540{\Mevc2}
\cite{Theta:LEPS,Theta:DIANA,Theta:CLASd,Theta:SAPHIR,Theta:CLASp,
Theta:NEUTRINO,Theta:HERMES,Theta:SVD,Theta:COSY,Theta:JINR,Theta:ZEUS,Theta:JINRBC}.
Figure \ref{fig:WA8902} shows a collection of the first nine
published results which gave evidence for the existence of the so
called {\t1540}. While the number of positive observations seems
to be quite convincing, when plotting the data points with error
bars but without background curves to guide the eye it becomes
obvious that the limited statistics is a common drawback of the
individual observations. It is remarkable that the event
statistics is nearly independent of the experimental situation and
it is disturbing that the peak positions differ significantly in
the various experiments. On the other hand at least 11 experiments
have reported negative search results
\cite{NTheta:HYPERCP,NTheta:BES,NTheta:HERAB,NTheta:SPHINX,NTheta:PHENIX,NTheta:BELLE,NTheta:LEP,NTheta:CDF,
NTheta:BABAR,NTheta:FOCUS,NTheta:LASS}. It was argued that this
discrepancy may be due to  very different production cross
sections in the various reaction processes (see e.g. Refs.
\cite{Theory:Karliner,Theory:Titov,Theory:Nussinov,Theory:Nam,Theory:Mart,Theory:Lipkin}).
Facing such a situation, further high statistics searches for this
resonance under different experimental conditions -- e.g.
different beam particles -- are highly desirable.

The hyperon beam experiment WA89 at CERN ran from 1990 to 1994 in
the West Hall. Its primary goal was the study of charmed particles
and their decays. At the same time it collected a high statistics
data sample of hyperons and hyperon resonances, among these \la\
decays, and also   \ks\ decays. We have already published the
negative result of a search for the pentaquark candidate
$\Phi(1860)$ , alternatively called $\Xi(1860)^{--}$
\cite{WA89:Ximm}. Here we report  a search for the pentaquark
candidate \t1540\ in the \ksp\ decay channel, produced inclusively
in  $\Sigma^-$-nucleus reactions. The results are based on the
data collected in the years 1993 and 1994.

The hyperon beamline selected negatively charged particles with a
mean momentum of 340 \gevc1 and a momentum spread of $\sigma
(p)/p=9\%$. At the experimental target, the $\pi^-$ to $\Sigma^-$
ratio of the beam was about 2.3. The beam pions were strongly
suppressed at the trigger level by a set of transition radiation
detectors resulting in a remaining pion contamination of about
12\%. In addition the beam contained small admixtures of $K^-$ and
$\Xi^-$. The experimental target itself consisted of one copper
slab with a thickness of 0.025 $\lambda_I $ in beam direction,
followed by three carbon (diamond powder) slabs of 0.008
$\lambda_I $ each. The trajectories of incoming and outgoing
particles were measured in silicon microstrip detectors upstream
and downstream of the targets. Only events with a reconstructed
interaction vertex in the targets and the surrounding counters
were retained in the analysis.

The momenta of charged particles were measured in a magnetic
spectrometer equipped with MWPCs and drift chambers. The
spectrometer magnet was placed with its center 13.6 m downstream
of the target, thus providing a field-free decay zone of about 10m
length for hyperons and $K^0_s$ emerging from the target.

A ring-imaging  Cherenkov counter placed downstream of the
spectrometer magnet provided particle identification. It was
followed by a leadglass electromagnetic calorimeter and an
iron/scintillator hadron calorimeter, which were not used in this
analysis.

\ks\ were reconstructed in the decay $K^0_s \rightarrow \pi^+\pi^-
$, using all pairs of positive and negative particles which formed
a decay vertex in the decay zone. $\Lambda \rightarrow p\pi^-$
decays with decay particle momenta corresponding to    $K^0_s
\rightarrow \pi^+\pi^- $ decays can produce a spurious mass peak
at 1540{\Mevc2}, if a mirror image of the decay proton is used in
the search for \ksp\ decays \cite{misha1,NTheta:HYPERCP}. To avoid
this, we excluded \ks\ candidates with a reconstructed $p\pi^-$
mass within $\pm 2 \, \sigma_m(\Lambda )$ of the \la\ mass
($\sigma_m(\Lambda )$ was 1.8 \Mevc2 at low momenta and 2.8 \Mevc2
at 200 {\gevc1}). This requirement reduced the \ks\ sample by 3\%
and the background by 1/3.

\begin{figure}[ht]
\centering
\includegraphics[width=9cm]{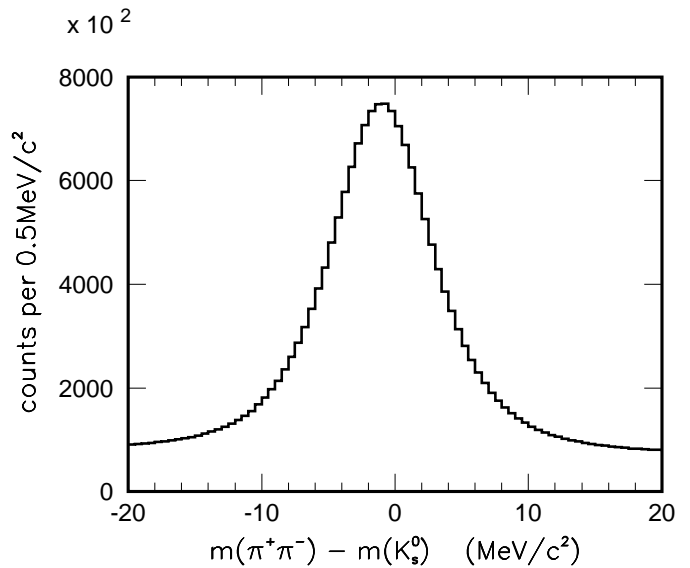}
\caption{Reconstructed mass distribution m($\pi^+\pi^-$) -
m$(K^0_s )$ of \ks\ candidates.} \label{fig:ks}
\end{figure}

The reconstructed  $\pi^+ \pi^-$ mass distribution of the
remaining \ks\ candidates is shown in fig. \ref{fig:ks}. The peak
from \ks\ decays contains about 13 million events, their momentum
spectrum extends from 10\gevc1 to about 200{\gevc1}. Above this
momentum, very few \ks\ are left, and they do not contribute to
\ksp\ effective masses below 1570{\Mevc2}. The mass resolution is
$\sigma_m(K^0_s )$ = 4\Mevc2 at low momenta and increases to
$\sigma_m(K^0_s )$ = 7\Mevc2 at 200{\gevc1}. Candidates with a
reconstructed $\pi^+\pi^-$ mass within $\pm 2\, \sigma_m$ of the
\ks\ mass were retained for further analysis.

All positive particles with a reconstructed track extending from
the microstrip counters downstream of the target to the wire
chambers beyond the spectrometer were considered as proton
candidates, excluding of course the $\pi^+$ from the \ks\ decay.
Requiring  track reconstruction in the microstrip
 counters rejected most of the protons from \la\ decays.
The track had to be inside the acceptance of the RICH counter,
which implies a momentum threshold at around 12{\gevc1}. Since the
proton threshold of the RICH was at 38 \gevc1 we did not require
proton identification, but rejected clearly identified $\pi^+$ and
$K^+$ (thresholds at 5.5 and 20{\gevc1}, resp.). From a study of
reconstructed \la\ decays, we determined that this requirement
rejected 4\% or less of genuine protons at all momenta, while the
\ksp\ candidate sample was reduced by a factor of 3.

\begin{figure}[ht]
\centering
\includegraphics[width=9cm]{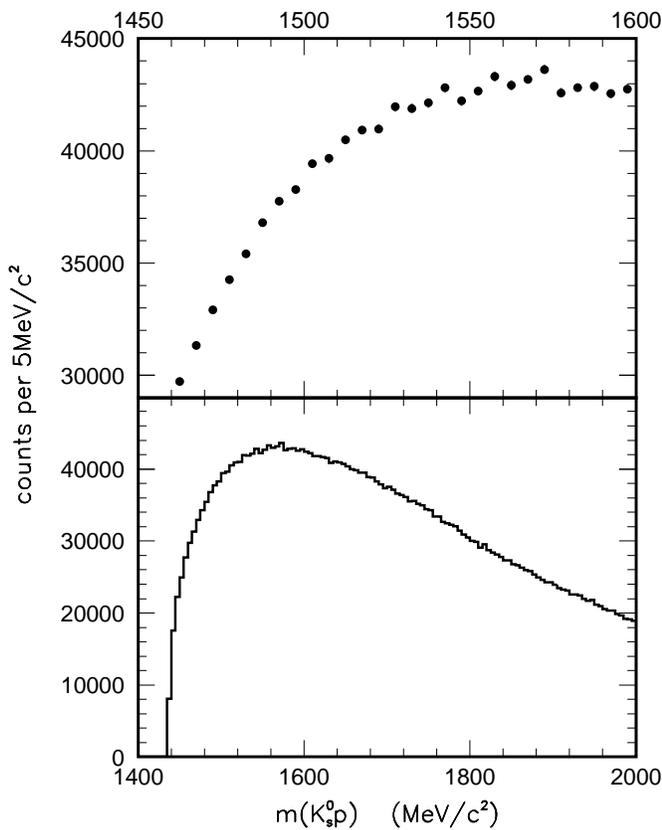}
\caption{Invariant mass spectrum of all observed $K^0_sp$
candidates. The upper plot shows an extended view of the region
around 1540{\Mevc2}. The statistical errors are approximately of
the size of the dots.} \label{fig:ksp_mass_all}
\end{figure}

\begin{figure}[ht]
\centering
\includegraphics[width=9cm]{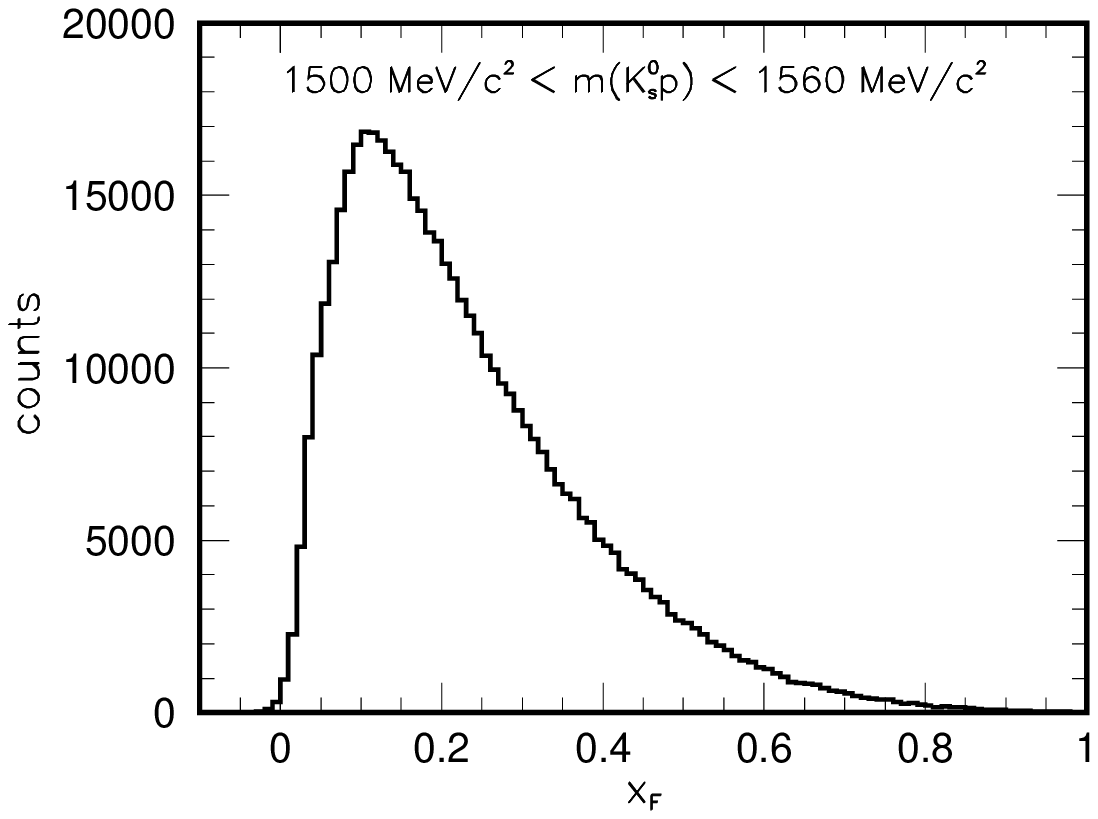}
\caption{\xf\ distribution of for $K^0_s p$ pairs in the mass
range 1500{\Mevc2}$<m(K^0_s p)<$1560{\Mevc2}.}
\label{fig:ksp_xf_thetaregion}
\end{figure}

The final \ksp\ sample contained 5.2 million \ksp\ candidates.
Fig. \ref{fig:ksp_mass_all} shows the \ksp\ mass distribution of
all candidates up to 2 GeV. No narrow signal is visible in this
plot, neither did we see narrow signals around an invariant mass
of 1540{\Mevc2} in subsamples of \xf\ or transverse momentum {\pt}
\cite{sigmastar}. We define \xf\ as $x_F=2p_L^*/\sqrt{s}$, where
$p_L^*$ is the \ksp\ momentum component in beam direction in the
beam-nucleon CMS and $\sqrt{s}$ is the invariant mass of the
beam-nucleon system. In our case, $\sqrt{s}$ = 25.2 GeV. The \xf\
distribution is shown in fig. \ref{fig:ksp_xf_thetaregion} for the
\ksp\ mass region between 1500 and 1560{\Mevc2}, it starts at
$x_F=0.05$ and thus covers part of the central production region.

Upper limits on the \tp\ production cross sections were calculated
separately for the copper and carbon targets, in bins of \xf\ as
listed in col. 1 of Table \ref{tab:results}. We used four mass
windows of 20\Mevc2 width, centered at 1520, 1530, 1540 and
1550{\Mevc2}, resp., for $i=1,2,3,4$, thus covering the full range
of reported values for the \tp\ mass. The width was chosen taking
into account our mass resolution, $\sigma_m(K^0_sp)$ = 4{\Mevc2},
and the reported values for the intrinsic  width of the \tp . The
observed number of  \ksp\ combinations in each mass window is
$n_i$. From a fit to the observed \ksp\ mass spectrum between 1460
and 1700{\Mevc2}
we calculated the expected non-resonant backgrounds $b_i$. Upper
limits  $n_{max}$ on the number  of $\Theta^+ \rightarrow K^0_sp$
decays were then obtained by the formula $n_{max}\, =\, max_{i} \{
max(0,n_i-b_i)\, +\, 3\sqrt{b_i} \} $ and are listed in columns 2
and 5 of Tab. \ref{tab:results}. These limits have a confidence
level of 99\% and scale approximately with the square root of the
width of the search window.

Upper limits on the product of $BR$, the \t1540\ $\rightarrow
K^0_sp$ decay branching ratio, and the differential production
cross sections $d\sigma_A /dx_F$ {\em per nucleus} are given in
columns 3 and 6 of Tab. \ref{tab:results}. Assuming the dependence
of the cross section on the mass number to be $\sigma_A \propto
\sigma_0\cdot A^{2/3}$, where $\sigma_0$ is the cross section {\em
per nucleon}, we finally obtained the limits on $BR\cdot d\sigma_0
/dx_F$ in columns 4 and 7  of the table.

\begin{table}[ht]
\centering
\begin{tabular}{|c||r|r|r||r|r|r|}
\hline
 & \multicolumn{3}{|c||}{copper target} &
\multicolumn{3}{|c|}{carbon target } \\ \cline{2-7}
  & & \multicolumn{2}{|c||}{ $BR\cdot d\sigma$/d$x_F$ [$\mu$b] }
  & & \multicolumn{2}{|c|}{ $BR\cdot d\sigma$/d$x_F$ [$\mu$b] } \\
 $x_F$ & $n_{max}$ & $\;\; d\sigma_A$& $d\sigma_0$ &
$n_{max}$ & $\;\; d\sigma_A$& $d\sigma_0$\\ \hline \hline
 0.05 - 0.15 & 520 & 230 & 14.5 & 550 & 105 & 20.0 \\ \hline
 0.15 - 0.25 & 500 & 205 & 13.0 & 480 &  80 & 15.3 \\ \hline
 0.25 - 0.35 & 340 & 140 &  8.8 & 350 &  55 & 10.5 \\ \hline
 0.35 - 0.45 & 390 & 140 &  8.8 & 290 &  40 &  7.6 \\ \hline
 0.45 - 0.55 & 250 & 65  & 4.1  & 240 & 25  &  4.8 \\ \hline
 0.55 - 0.65 & 190 & 53  & 3.3  & 160 & 16  &  3.0  \\ \hline
 0.65 - 0.75 & 115 & 33  & 2.1  & 130 & 13  &  2.5 \\ \hline
 0.75 - 0.85 & 70  & 21  & 1.3  &  55 &  6  &  1.1 \\ \hline
  $>$ 0.85   & 35  & 11  & 0.7  &  45 &  5  &  1.0 \\ \hline
 & & $BR\cdot \sigma_A$ & $BR\cdot \sigma_0$ &  &
 $BR\cdot \sigma_A$ & $BR\cdot \sigma_0$ \\
\cline{3-4} \cline{6-7} & & 38 & 2.4 & & 15 & 2.9 \\ \hline
\end{tabular}
\caption{Upper limits on yields and cross sections. BR denotes the
{\t1540}$\rightarrow K^0_sp$ decay branching ratio. $\sigma_A$ and
$\sigma_0$ denote cross sections per nucleus and per nucleon,
respectively.} \label{tab:results}
\end{table}

Limits on the integrated production cross sections $\sigma$ were
calculated by summing quadratically the contributions $( d\sigma
/dx_F ) \cdot \Delta x_F$ in the nine individual $x_F$ bins. The
results are $BR\cdot \sigma_A (x_F>0.05)<$ 38  and $< $15 $\mu$b
per nucleus for the copper and carbon target, respectively. An
extrapolation to the cross sections per nucleon yields the two
values $BR\cdot\sigma_0 (x_F>0.05)<$ 2.4 and $<$ 2.9 $\mu$b per
nucleon. Since these are statistically independent upper limits,
we can combine them to obtain $BR\cdot \sigma_0 <$ 1.8 $\mu b$ per
nucleon for \t1540\ production by $\Sigma^-$ of 340 \gevc1 in the
region $x_F>$ 0.05.

For a comparison of our result to observations of or searches for
the \t1540\ we concentrate on hadronic reactions. It is
interesting to note that all these experiments investigated the
\ksp\ decay channel, but only the SPHINX experiment searched in
the $K^+n$ decay channel as well. Four experiments have reported
observations of the \t1540\
\cite{Theta:COSY,Theta:JINR,Theta:SVD,Theta:DIANA}. The COSY-TOF
collaboration using a proton beam of 2.95 \gevc1 and a liquid
hydrogen target, has measured a cross section $\sigma_0$ = 0.4
$\mu$b per nucleon for {\sl exclusive} production in the reaction
$pp \rightarrow \Sigma^+(K^0p)$ \cite{Theta:COSY}. This value is
below our upper limit, but an exclusive production cross section
that close to the reaction threshold cannot be compared to
inclusive production cross sections at energies of several hundred
GeV. The JINR propane bubble chamber group using a proton beam of
10 \gevc1 has measured a total production cross section
$\sigma_{propane}$ = 90 $\mu$b \cite{Theta:JINR}. Again assuming a
dependence of the cross section on the mass number as $\sigma_A
\propto \sigma_0\cdot A^{2/3}$, one obtains a production cross
section $\sigma_0$ = 3.8 $\mu$b per nucleon, which is larger by a
factor of 2 than our limit. The SVD Collaboration using a proton
beam of 70 \gevc1 and a combined carbon, silicon and lead target,
has measured  a production cross section $\sigma_0$ = 30 - 120
$\mu$b per nucleon for $x_F>0$ \cite{Theta:SVD}. This is much
higher than our upper limit in practically the same kinematic
range. The DIANA collaboration using a $K^+$ beam of 0.85 \gevc1
and a Xenon bubble chamber has not measured a cross section
\cite{Theta:DIANA}.

Negative search results were reported from at least 11 experiments
\cite{NTheta:BES,NTheta:HYPERCP,NTheta:HERAB,NTheta:SPHINX,NTheta:PHENIX,NTheta:BELLE,NTheta:CDF,NTheta:LEP,
NTheta:BABAR,NTheta:FOCUS,NTheta:LASS}. Out of these 6 experiments
studied hadronic induced interactions
\cite{NTheta:HYPERCP,NTheta:HERAB,NTheta:SPHINX,NTheta:PHENIX,NTheta:BELLE,NTheta:CDF}
Usually these collaborations have compared their \tp\ production
limits with their $\Lambda(1520)$ observations, and have obtained
limits below 3 \% on the event or production ratio of \t1540\
w.r.t. $\Lambda(1520)$. This we cannot do, although we do observe
$\Lambda(1520)$ decays, because in our experiment two-body decay
channels were suppressed in  the trigger. We can, however, compare
our result to the HERA-B result of $BR\cdot d\sigma /dy <$ 4 - 16
$\mu$b per nucleon at 95\% CL for \tp\ masses between 1521 and
1555 {\Mevc2}, at rapidity $y_{cm}\approx 0$. This value
corresponds to $BR\cdot d\sigma /dx_F <$ 30 - 120 $\mu$b per
nucleon, to be compared to our result $BR\cdot d\sigma /dx_F <$ 12
$\mu$b at 99\% CL and for $0.05<x_F<0.15$ (this limit was obtained
by combining the statistically independent carbon and copper
target results).

If the \t1540\ exists, as many experiments suggest, then the cross
sections for \tp\ production in hadronic reactions at higher
energies are surprisingly low compared to the production of
hyperon resonances. This fact by itself could provide important
information on the nature of the \t1540 .


\begin{thebibliography}{090}
\bibitem{sigmastar}
At large x$_F >0.8$ we do however observe a broad ($\Gamma\simeq$
90{\Mevc2}) resonance like structure  at a mass of $\simeq$
1750{\Mevc2} which is possibly related to known $\Sigma^*$
resonances. A detailed analysis of this structure will be
presented in a future paper.
\end{thebibliography}
\end{document}